\documentclass[11pt]{article}
\usepackage{latexsym}
\usepackage{amssymb}
\usepackage{graphics,graphicx}
\usepackage{subfig}
\usepackage{tabularx}
\usepackage{amsfonts}
\usepackage{amsmath}
\usepackage{enumerate}
\usepackage[dvips]{color}

\newcommand{\footmsg}[1]{%
  \let\temp\thempfn%
  \def\thempfs{}
  \footnotetext{#1}
  \let\tempfn\temp}

\definecolor{purple}{RGB}{143,52,43}

\begin{document}

\newcommand{\singlespace} {\baselineskip=12pt
\lineskiplimit=0pt \lineskip=0pt }
\def\ds{\displaystyle}

\newcommand{\beq}{\begin{equation}}
\newcommand{\eeq}{\end{equation}}
\newcommand{\lb}{\label}
\newcommand{\beqar}{\begin{eqnarray}}
\newcommand{\eeqar}{\end{eqnarray}}
\newcommand{\barr}{\begin{array}}
\newcommand{\earr}{\end{array}}

\newcommand{\jump}{\parallel}

\def\c{{\circ}}

\newcommand{\Ehat}{\hat{E}}
\newcommand{\That}{\hat{\bf T}}
\newcommand{\Ahat}{\hat{A}}
\newcommand{\chat}{\hat{c}}
\newcommand{\shat}{\hat{s}}
\newcommand{\khat}{\hat{k}}
\newcommand{\muhat}{\hat{\mu}}
\newcommand{\mc}{M^{\scriptscriptstyle C}}
\newcommand{\mei}{M^{\scriptscriptstyle M,EI}}
\newcommand{\mec}{M^{\scriptscriptstyle M,EC}}

\newcommand{\hbeta}{{\hat{\beta}}}
\newcommand{\rec}[2]{\left( #1 #2 \ds{\frac{1}{#1}}\right)}
\newcommand{\rep}[2]{\left( {#1}^2 #2 \ds{\frac{1}{{#1}^2}}\right)}
\newcommand{\derp}[2]{\ds{\frac {\partial #1}{\partial #2}}}
\newcommand{\derpn}[3]{\ds{\frac {\partial^{#3}#1}{\partial #2^{#3}}}}
\newcommand{\dert}[2]{\ds{\frac {d #1}{d #2}}}
\newcommand{\dertn}[3]{\ds{\frac {d^{#3} #1}{d #2^{#3}}}}

\def\bob{{\, \underline{\overline{\otimes}} \,}}

\def\ob{{\, \underline{\otimes} \,}}
\def\scalp{\mbox{\boldmath$\, \cdot \, $}}
\def\gdp{\makebox{\raisebox{-.215ex}{$\Box$}\hspace{-.778em}$\times$}}

\def\daa{\makebox{\raisebox{-.050ex}{$-$}\hspace{-.550em}$: ~$}}

\def\mK{\mbox{${\mathcal{K}}$}}
\def\cK{\mbox{${\mathbb {K}}$}}

\def\Xint#1{\mathchoice
   {\XXint\displaystyle\textstyle{#1}}%
   {\XXint\textstyle\scriptstyle{#1}}%
   {\XXint\scriptstyle\scriptscriptstyle{#1}}%
   {\XXint\scriptscriptstyle\scriptscriptstyle{#1}}%
   \!\int}
\def\XXint#1#2#3{{\setbox0=\hbox{$#1{#2#3}{\int}$}
     \vcenter{\hbox{$#2#3$}}\kern-.5\wd0}}
\def\ddashint{\Xint=}
\def\fpint{\Xint=}
\def\dashint{\Xint-}
\def\cpvint{\Xint-}
\def\intl{\int\limits}
\def\cpvintl{\cpvint\limits}
\def\fpintl{\fpint\limits}
\def\ointl{\oint\limits}

\def\half{{\scriptstyle{\frac{1}{2}}}}

\def\bA{{\bf A}}
\def\ba{{\bf a}}
\def\bB{{\bf B}}
\def\bb{{\bf b}}
\def\bc{{\bf c}}
\def\bC{{\bf C}}
\def\bD{{\bf D}}
\def\bE{{\bf E}}
\def\be{{\bf e}}
\def\bbf{{\bf f}}
\def\bF{{\bf F}}
\def\bG{{\bf G}}
\def\bg{{\bf g}}
\def\bi{{\bf i}}
\def\bH{{\bf H}}
\def\bK{{\bf K}}
\def\bL{{\bf L}}
\def\bM{{\bf M}}
\def\bN{{\bf N}}
\def\bn{{\bf n}}
\def\bm{{\bf m}}
\def\b0{{\bf 0}}
\def\bo{{\bf o}}
\def\bX{{\bf X}}
\def\bx{{\bf x}}
\def\bP{{\bf P}}
\def\bp{{\bf p}}
\def\bQ{{\bf Q}}
\def\bq{{\bf q}}
\def\bR{{\bf R}}
\def\bS{{\bf S}}
\def\bs{{\bf s}}
\def\bT{{\bf T}}
\def\bt{{\bf t}}
\def\bU{{\bf U}}
\def\bu{{\bf u}}
\def\bv{{\bf v}}
\def\bw{{\bf w}}
\def\bW{{\bf W}}
\def\by{{\bf y}}
\def\bz{{\bf z}}

\def\T{{\bf T}}
\def\Te{\textrm{T}}

\def\e{{\rm{e}}}
\def\Id{{\bf I}}
\def\p{{\rm{p}}}
\def\t{{\rm{t}}}
\def\bxi{\mbox{\boldmath${\xi}$}}
\def\balpha{\mbox{\boldmath${\alpha}$}}
\def\bbeta{\mbox{\boldmath${\beta}$}}
\def\bepsilon{\mbox{\boldmath${\epsilon}$}}
\def\bvarepsilon{\mbox{\boldmath${\varepsilon}$}}
\def\bomega{\mbox{\boldmath${\omega}$}}
\def\bphi{\mbox{\boldmath${\phi}$}}
\def\bsigma{\mbox{\boldmath${\sigma}$}}
\def\bfeta{\mbox{\boldmath${\eta}$}}
\def\bDelta{\mbox{\boldmath${\Delta}$}}
\def\btau{\mbox{\boldmath $\tau$}}

\def\tr{{\rm tr}}
\def\dev{{\rm dev}}
\def\div{{\rm div}}
\def\Div{{\rm Div}}
\def\Grad{{\rm Grad}}
\def\grad{{\rm grad}}
\def\Lin{{\rm Lin}}
\def\Sym{{\rm Sym}}
\def\Skw{{\rm Skew}}
\def\abs{{\rm abs}}
\def\Re{{\rm Re}}
\def\Im{{\rm Im}}
\def\sign{{\rm sign}}

\def\capB{\mbox{\boldmath${\mathsf B}$}}
\def\capC{\mbox{\boldmath${\mathsf C}$}}
\def\capD{\mbox{\boldmath${\mathsf D}$}}
\def\capE{\mbox{\boldmath${\mathsf E}$}}
\def\capG{\mbox{\boldmath${\mathsf G}$}}
\def\tcapG{\tilde{\capG}}
\def\capH{\mbox{\boldmath${\mathsf H}$}}
\def\capK{\mbox{\boldmath${\mathsf K}$}}
\def\capL{\mbox{\boldmath${\mathsf L}$}}
\def\capM{\mbox{\boldmath${\mathsf M}$}}
\def\capR{\mbox{\boldmath${\mathsf R}$}}
\def\capW{\mbox{\boldmath${\mathsf W}$}}

\def\i{\mbox{${\mathrm i}$}}

\def\mC{\mbox{\boldmath${\mathcal C}$}}

\def\d{\mbox{{d}}}

\def\mB{\mbox{${\mathcal B}$}}
\def\mE{\mbox{${\mathcal{E}}$}}
\def\mL{\mbox{${\mathcal{L}}$}}
\def\mK{\mbox{${\mathcal{K}}$}}
\def\mV{\mbox{${\mathcal{V}}$}}

\def\C{\mbox{\boldmath${\mathcal C}$}}
\def\E{\mbox{\boldmath${\mathcal E}$}}

\def\ACME{{ Arch. Comput. Meth. Engng.\ }}
\def\ARMA{{ Arch. Rat. Mech. Analysis\ }}
\def\AMR{{ Appl. Mech. Rev.\ }}
\def\ASCEEM{{ ASCE J. Eng. Mech.\ }}
\def\acta{{ Acta Mater. \ }}
\def\AMM{{ Acta Metall. Mater. \ }}
\def\CMAME {{ Comput. Meth. Appl. Mech. Engrg.\ }}
\def\CRAS{{ C. R. Acad. Sci., Paris\ }}
\def\EFM{{ Eng. Fracture Mechanics\ }}
\def\EJMA{{ Eur.~J.~Mechanics-A/Solids\ }}
\def\IJES{{ Int. J. Eng. Sci.\ }}
\def\IJF{Int. J. Fracture}
\def\IJMS{{ Int. J. Mech. Sci.\ }}
\def\IJNAMG{{ Int. J. Numer. Anal. Meth. Geomech.\ }}
\def\IJP{{ Int. J. Plasticity\ }}
\def\IJSS{{ Int. J. Solids Structures\ }}
\def\IngA{{ Ing. Archiv\ }}
\def\JAM{{ J. Appl. Mech.\ }}
\def\JAP{{ J. Appl. Phys.\ }}
\def\JEM{{J. Engrg. Mech., ASCE\ }}
\def\JE{J. Elasticity\ }
\def\JM{{ J. de M\'ecanique\ }}
\def\JMPS{{ J. Mech. Phys. Solids\ }}
\def\Macro{{ Macromolecules\ }}
\def\MOM{{ Mech. Materials\ }}
\def\MMS{{ Math. Mech. Solids\ }}
\def\MMT{{ Metall. Mater. Trans. A}}
\def\MPCPS{{ Math. Proc. Camb. Phil. Soc.\ }}
\def\MRC{{ Mec. Res. Comm. \ }}
\def\MSE{{ Mater. Sci. Eng.}}
\def\PMPS{{ Proc. Math. Phys. Soc.\ }}
\def\PRE{{ Phys. Rev. E\ }}
\def\PRSL{{ Proc. R. Soc.\ }}
\def\rock{{ Rock Mech. and Rock Eng.\ }}
\def\QAM{{ Quart. Appl. Math.\ }}
\def\QJMAM{{ Quart. J. Mech. Appl. Math.\ }}
\def\SCRMAT{{ Scripta Mater.\ }}
\def\SM{{\it Scripta Metall. }}


\def\salto#1#2{
\left[\mbox{\hspace{-#1em}}\left[#2\right]\mbox{\hspace{-#1em}}\right]}

\def\medio#1#2{
\mbox{\hspace{-#1em}}<#2>\mbox{\hspace{-#1em}}}




\title{Eshelby-like forces acting on elastic structures:
\\ theoretical and experimental proof}

\author{D. Bigoni$^0$, F. Dal Corso, F. Bosi and D. Misseroni\\
\normalsize{University of Trento, via Mesiano 77, I-38123 Trento, Italy} \\
\normalsize{e-mail: davide.bigoni@unitn.it; francesco.dalcorso@unitn.it;}\\
\normalsize{federico.bosi@unitn.it; diego.misseroni@unitn.it}
}
\date{}
\maketitle

\footnotetext[0]{Corresponding author: fax +390461282599; tel. +390461282507; e-mail davide.bigoni@unitn.it}

\begin{center}
\emph{Dedicated to Prof. Alain Molinari}
\end{center}
\vspace*{5mm}

\begin{abstract}
\noindent
The Eshelbian (or configurational) force is the main concept of a celebrated theoretical framework
associated with the motion of dislocations and, more in general, defects in solids. In a similar vein,
in an elastic structure where a (smooth and bilateral) constraint can move and release energy,
a force driving the configuration is generated, which therefore is called by analogy \lq Eshelby-like'
or \lq configurational'. This force (generated by a specific movable constraint) is derived both
via variational calculus and, independently, through an asymptotic approach. Its action on the elastic
structure is counterintuitive, but is fully substantiated and experimentally measured on a model structure
that has been designed, realized and tested.
These findings open a totally new perspective in the mechanics of deformable mechanisms, with possible broad applications, even at the nanoscale.
\end{abstract}


\section{Introduction}

Configurational (or: \lq material', \lq driving', \lq non-Newtonian') forces have been introduced by Eshelby (1951; 1956; 1970; 1975)
to describe the fact that massless (for instance: voids, microcracks, vacancies, or dislocations) or heavy (for instance inclusions)
defects may move within a solid body as a result of mechanical or thermal loading. The Eshelbian force is
defined as the negative gradient of the total potential energy $\mathcal{V}$ of a body with respect to the parameter
 $\kappa$ determining the configuration of the defect, namely,
$
-\partial \mathcal{V}(\kappa) /\partial\kappa$.

Examples are the crack-extension force of fracture mechanics, the Peach--Koehler force of dislocations, or the material force
 developing on a phase boundary in a solid under loading.
Nowadays configurational forces are the cornerstone of a well-developed theory (see for instance the monographs by Gurtin, 2000,
Kienzler and Herrmann, 2000,  and
Maugin, 1993, 2011,
and the journal special issues by Dascalu et al., 2010, and Bigoni and Deseri, 2011).

Let us consider an elastic structure in equilibrium upon load and assume that a (frictionless and bilateral) constraint can move --a feature which may be considered as a \lq defect'-- in a way to allow the system to reconfigure through a release of elastic energy, then a force  is generated, similar to an Eshelbian or configurational\footnote{
\lq Configurational force' is not to be confused with the follower forces analyzed for instance by Bigoni and Noselli, (2011), or with the tensile buckling analyzed by Zaccaria et al. (2011).
}
one.

To reveal this force in an indisputable way, and directly measure it,
the simple elastic structure sketched in Fig. \ref{system} has been designed, which inflection
length can change through sliding along a sleeve and therefore discloses (in two different
and independent ways, namely, using variational and asymptotic approaches) the presence of an Eshelby-like force.
The structure has been subsequently realized and instrumented (see Fig. \ref{fiorellino},
reporting a series of photos demonstrating the action of the Eshelby-like force),
so that the configurational force has
been measured at equilibrium and it is shown to perfectly match the theoretical predictions.
\begin{figure}[!htcb]
  \begin{center}
      \includegraphics[width= 8 cm]{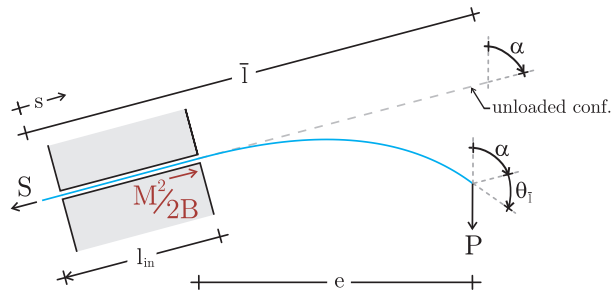}
\caption{\footnotesize Structural scheme of the elastic system employed to disclose
a Eshelby-like force. The elastic rod of total length $\bar{l}$ is subject to a
dead vertical load $P$ on its right end, is constrained with a sliding sleeve inclined at an angle $\alpha$
(with respect to the vertical) and has a axial dead force $S$ applied at its left end. The presence of the Eshelby-like force $M^2/(2B)$ influences the force $S$ at equilibrium, which results different from $P \cos\alpha$.}
\lb{system}
  \end{center}
\end{figure}

In this example configurational forces are non-zero, but small for small deflections\footnote{
The fact that these forces are small for small displacement does not mean that they are
always negligible, since their action is in a particular direction, which may be \lq unexpected'.
For instance, in the case of null axial dead load, $S=0$, and sliding sleeve
orthogonal to the vertical dead load $P$, $\alpha = \pi/2$ (Fig. \ref{system}),
the Eshelby-like force is the only axial action, so that equilibrium becomes impossible.
} and become progressively important when displacements grow. Their effects are counterintuitive and unexpected, so that for instance, the structure
shown in Fig. \ref{fiorellino}, which can (wrongly!) be thought to be unable to provide any axial action, is instead subject to an axial Eshelby-like force
transmitted by the sliding sleeve.
\begin{figure}[!htcb]
  \begin{center}
      \includegraphics[width= 12 cm]{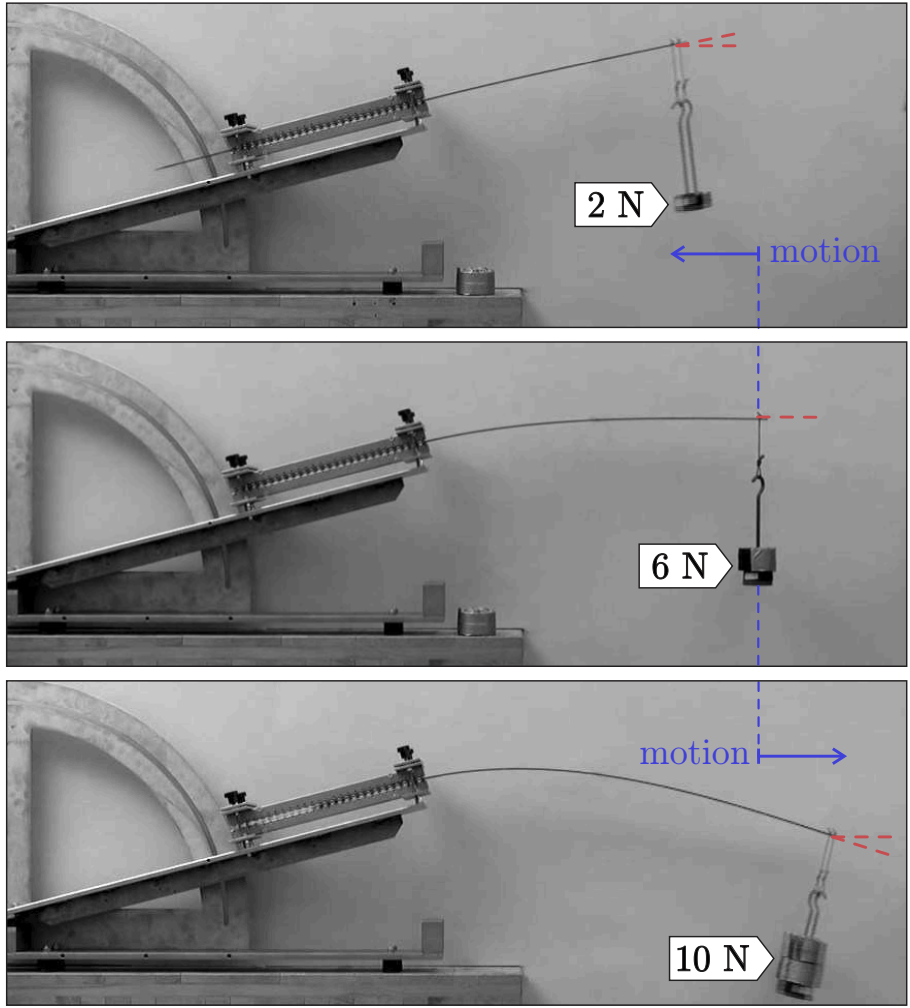}
\caption{\footnotesize The practical realization of the elastic structure shown in Fig. \ref{system} reveals an axial Eshelby-like force,
so that, while at low vertical force (2 N) the elastic rod tends, as expected, to
slip inside the sliding sleeve (upper photo), at 6 N the equilibrium is surprisingly possible
(note that the tangent at the loaded end of the elastic rod is horizontal, see the photo in the centre) and at
10 N the elastic rod is expelled from the sliding sleeve (lower photo),
even if the system is inclined at 15$^\circ$ with respect to the horizontal ($\alpha = 75^\circ$).}
\lb{fiorellino}
  \end{center}
\end{figure}
In particular, at the end of the sliding sleeve, the axial force $S$ at equilibrium with a load $P$ (inclined of $\alpha$ with respect to the rod's axis)
is not simply equal to $-P \cos \alpha$, as when the sliding sleeve is replaced by a movable clamp, but
will be determined (Section \ref{elasticazza}) to be a function of the rotation of the rod at its end, $\theta_{\bar{l}}$, as
\beq
\lb{bigonata}
S = - P\cos \left(\alpha + \theta_{\bar{l}} \right) = -P\cos \alpha +
{\color{purple}\underbrace{2P \left( \sin^2\frac{\theta_{\bar{l}}+\alpha}{2}-\sin^2\frac{\alpha}{2}\right)}_{Eshelby-like\,\,force} },
\eeq
which for for small deflections ($\sin \theta_{\bar{l}} \approx \theta_{\bar{l}}$) becomes
\beq
\lb{bigonata_small}
S = - P \cos \alpha + {\color{purple}\underbrace{P \frac{3\,v_{\bar{l}}}{2(\bar{l}-l_{in})}\sin\alpha}_{Eshelby-like\,\,force} },
\eeq
where $v_{\bar{l}}$ is the transversal displacement at the loaded end of the rod of length $\bar{l}-l_{in}$ (external to the sliding sleeve).
Eqs (\ref{bigonata}) and (\ref{bigonata_small}) show that there is an \lq unexpected' term (null if the elastic rod is constrained by a movable clamp instead of a sliding sleeve), defined as the \lq Eshelby-like force'.
Although there is a little abuse of notation\footnote{
The introduction of the nomenclature \lq Eshelby-like force' allows to distinguish
terms generated by the possibility of configurational changes of the system, while \lq Eshelby forces' must always vanish at equilibrium.
}, this definition is motivated by the fact that
the Eshelby-like force is null, would the total potential energy of the system be independent of a
configurational parameter.

The findings presented in this article demonstrate that movable constraints applied to elastic structures can generate configurational
forces and that these become dominant when deformations are sufficiently large. Configurational forces can be employed in the design of new deformable systems with challenging characteristics, which may find applications even at the micro- and nano-scale,
for instance, to control growth of a structural element.

\section{Eshelby-like force produced by a sliding sleeve}

An inextensible elastic rod (straight in its unloaded configuration, with bending stiffness $B$ and total length $\bar{l}$)
has one end constrained with a sliding sleeve, is subject to an edge axial (dead) force $S$,
and has the other end subject to a dead transversal load $P$ (inclined at an angle $\alpha$, see Fig. \ref{system}).
Introducing the curvilinear coordinate $s\in [0, \bar{l}]$, the length  $l_{in}$ of the segment of the rod internal
to a (frictionless, perfectly smooth and bilateral) sliding sleeve,
and the rotation $\theta(s)$ of the rod's axis, it follows that $\theta(s)=0$ for $s\in[0,l_{in}]$. Denoting by a
prime the derivative with respect to $s$, the bending moment along the elastic rod is
$M(s) = B\theta'(s)$, so that
at the loaded edge of the
rod, we have the zero-moment boundary condition $\theta '(\bar{l})=0$.

The total potential energy of the system is
\begin{equation}
\label{eq:EPT}
\mathcal{V}(\theta(s),l_{in})= B \intop_{l_{in}}^{\bar{l}}\frac{\left[\theta^{'}\left(s\right)\right]^{2}}{2}\d s
-P\left[\bar{l}\cos\alpha-\cos\alpha\intop_{l_{in}}^{\bar{l}}\cos\theta(s)\d s
+\sin\alpha\intop_{l_{in}}^{\bar{l}}\sin\theta(s)\d s\right] - S \, l_{in},
\end{equation}
which at equilibrium becomes
\begin{equation}
\label{equilib}\begin{array}{lll}
\mathcal{V}(\theta_{eq}(s, l_{eq}),l_{eq})=  & \ds B \intop_{l_{eq}}^{\bar{l}}\frac{\left[\theta^{'}_{eq}\left(s, l_{eq}\right)\right]^{2}}{2}\d s
-P\left[\bar{l}\cos\alpha-\cos\alpha\intop_{l_{eq}}^{\bar{l}}\cos\theta_{eq}(s, l_{eq})\d s \right.\\[5mm]
&\ds \left.+\sin\alpha\intop_{l_{eq}}^{\bar{l}}\sin\theta_{eq}(s, l_{eq})\d s\right] - S \, l_{eq},
\end{array}
\end{equation}
where $l_{eq}$ is the length of the elastic rod inside the sliding sleeve and $\theta_{eq}$
the rotation of the rod's axis at the equilibrium configuration.

The Eshelbian force related to the sliding in the sleeve can be calculated by taking the derivative with
respect to $l_{eq}$ of the total potential energy at equilibrium, eqn (\ref{equilib}). In particular, keeping into
account integration by parts
\beq
\theta_{eq}' \frac{\partial \theta_{eq}'}{\partial l_{eq}} = \left( \theta_{eq}'
\frac{\partial \theta_{eq}}{\partial l_{eq}} \right)' - \theta_{eq}'' \frac{\partial \theta_{eq}}{\partial l_{eq}} ,
\eeq
the equilibrium of the elastica
\beq
\lb{system_2}
\ds  B \theta_{eq}'' (s) + P \left[\cos\alpha\sin\theta_{eq}(s)+\sin\alpha\cos\theta_{eq}(s)\right]=0,\qquad s\in[l_{eq},\bar{l}]
\eeq
and the boundary condition $\theta_{eq}'(\bar{l})=0$, we arrive at the following expression for the Eshelby force
\beq
\lb{bellafica}
-\frac{\partial \mathcal{V}(l_{eq})}{\partial l_{eq}} = B\frac{[\theta_{eq}'(l_{eq})]^2}{2} +
B \theta_{eq}'(l_{eq}) \left.\frac{\partial \theta_{eq}}{\partial l_{eq}}\right|_{s=l_{eq}} + P \cos\alpha + S.
\eeq
The fact that $\theta_{eq}$ is a function of $s-l_{eq}$ and of the angle of rotation of the beam
at the loaded end $\theta_{\bar{l}}$ (function itself of $l_{eq}$), but is always zero at $s=l_{eq}$ for all $\theta_{\bar{l}}$, yields
\beq
\left. \frac{\partial \theta_{eq}}{\partial l_{eq}}\right|_{s=l_{eq}} = - \theta_{eq}'(l_{eq}),
\eeq
so that the vanishing of the derivative
with respect to $l_{eq}$ of the total potential energy, eqn (\ref{bellafica}), represents the axial equilibrium
\begin{equation}
\label{nonnoeshelby}
{\color{purple}\underbrace{\frac{M^2}{2B}}_{Eshelby-like\,\,force} }= S+P\cos\alpha ,
\end{equation}
where $M = B\theta_{eq}'(l_{eq})$ is the reaction moment, equal to $P e$, where $e$ is the load eccentricity (to the sliding sleeve).

Although the Eshelby force must vanish at equilibrium,
the contribution $M^2/(2B)$ is a \lq counterintuitive term' which depends on the configurational parameter $l_{eq}$ (and would be absent if the
elastic rod would be constrained with a movable clamp instead than a sliding sleeve) and is for this reason indicated as the \lq Eshelby-like force'.

This term has wrongly been neglected by a number of authors
who have considered sliding sleeve constraints,
while a term $M^2/(2B)$ correctly enters in calculations referred in a different context, namely, adhesion mechanics,
in which it is equated to an \lq adhesion energy' (Majidi, 2007; Majidi et al. 2012).

Since equilibrium is only possible when eqn (\ref{nonnoeshelby}) is satisfied,
the presence of the Eshelby-like force (parallel to the direction of sliding) explains the reason why the
equilibrium is possible for the configuration shown in the central photo in Fig. \ref{fiorellino} and why
the rod is \lq expelled' from the sliding sleeve in the lower photo.

In the next sections the existence of the Eshelby-like force (\ref{nonnoeshelby})
will be demonstrated via two different and independent approaches (an asymptotic method and a variational technique).

\subsection{Asymptotic approach}\lb{asint}

The Eshelbian force (\ref{nonnoeshelby}) can be obtained via an asymptotic approach.
This has been found in a forgotten article published in Russian by Balabukh et al. (1970). The main idea is to consider
an imperfect sliding sleeve (Fig. \ref{configurational}) having a small gap $\Delta$ (the distance between the two rigid, frictionless and parallel
surfaces realizing the sliding device),
so that the perfect sliding sleeve case is recovered when the gap is null, $\Delta=0$. Within this space,
the elastic rod is deflected, so that $\vartheta(\Delta)$ denotes the angle at its right contact point,
where the forces $H$, $V$, $M$ are applied.
The length of the rod
detached from the two surfaces representing the imperfect sliding sleeve is denoted with $a(\Delta)$.
The frictionless contact generates the
reaction forces $R$ and $Q$, in equilibrium with the axial dead force $S$ at the other end.
For small $\Delta$, the equilibrium is given by
\beq
Q=\frac{M}{a(\Delta)},\qquad
R=V+\frac{M}{a(\Delta)},\qquad
S=\left(V+\frac{M}{a(\Delta)}\right)\vartheta(\Delta)-H.
\eeq
On application of the virtual work for a linear elastic inextensible rod yields
the geometric quantities $a(\Delta)$ and $\vartheta(\Delta)$
\beq
a(\Delta)=\sqrt{\frac{6B \Delta}{M}},\qquad
\vartheta(\Delta)=\frac{1}{2}\sqrt{\frac{6M \Delta}{B}},
\eeq
so that forces $Q$, $R$ and $S$ can be written as
\beq\lb{reaction_final}
Q=M\sqrt{\frac{M}{6B \Delta}},\qquad
R=V+M\sqrt{\frac{M}{6B \Delta}},\qquad
S=\frac{M^2}{2B}+\frac{V}{2}\sqrt{\frac{6M\Delta}{B }}-H.
\eeq
In the limit of perfect (zero-thickness) sliding sleeve, $\Delta\rightarrow0$,
the horizontal component of the reaction $R$ {\it does not vanish, but} becomes the Eshelbian force (\ref{nonnoeshelby})
\beq
\lim_{\Delta\rightarrow0}R(\Delta) \vartheta(\Delta)={\color{purple}\frac{M^2}{2 B}}.
\eeq

\begin{figure}[!htcb]
  \begin{center}
      \includegraphics[width= 8 cm]{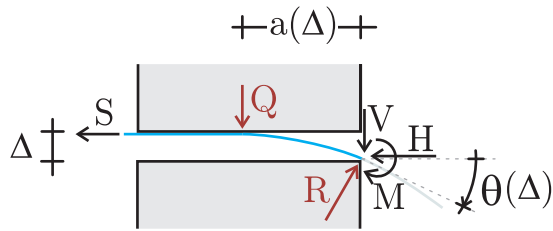}
\caption{\footnotesize Deformed configuration of an elastic rod within an imperfect sliding sleeve made up of two smooth, rigid and frictionless planes placed at a distance $\Delta$.
Applied and reaction forces provide in the limit $\Delta\rightarrow0$ the Eshelby-like force.}
\lb{configurational}
  \end{center}
\end{figure}

\subsection{Variational approach}

The total potential energy (\ref{eq:EPT}) has a movable boundary $l_{in}$, so that it is expedient
(Courant and Hilbert, 1953, see also Majidi et al. 2012)
to introduce a small parameter $\epsilon$ and take variations (subscript \lq $var$') of an equilibrium configuration (subscript \lq $eq$') in the form
\begin{equation}
\begin{array}{cc}
\theta(s,\epsilon)=\theta_{eq}(s)+\epsilon\theta_{var}(s),\quad\quad & l_{in}(\epsilon)=l_{eq}+\epsilon l_{var}\end{array},
\end{equation}
with the boundary conditions
\begin{equation}
\theta_{eq}(l_{eq})=0, ~~
\theta(l_{eq}+\epsilon l_{var})=0, ~~
\theta_{eq}^{'}(\bar{l})=0.
\label{eq:bc}
\end{equation}
A Taylor series expansion of $\theta(l_{in})$ for small $\epsilon$  yields
\begin{equation}
\theta(l_{eq}+\epsilon l_{var},\epsilon) = \theta_{eq}(l_{eq})+\epsilon\left(\theta_{var}(l_{eq})+\theta_{eq}^{'}(l_{eq})l_{var}\right)
+\frac{\epsilon^{2}}{2}l_{var}\left(2\theta_{var}^{'}(l_{eq})+\theta_{eq}^{''}(l_{eq})l_{var}\right)+\mathcal{O}\left(\epsilon^{3}\right),
\label{eq:vartot}
\end{equation}
so that the boundary conditions (\ref{eq:bc}) lead to
the following compatibility equations
\begin{equation}
\theta_{var}(l_{eq})+\theta_{eq}^{'}(l_{eq})l_{var}=0, ~~
2\theta_{var}^{'}(l_{eq})+\theta_{eq}^{''}(l_{eq})l_{var}=0.
\label{eq:compatcond}
\end{equation}

Taking into account the Leibniz rule of differentiation and the boundary (\ref{eq:bc}) and compatibility (\ref{eq:compatcond}) conditions, through
integration by parts, the first variation of the functional $\mathcal{V}$ is
\begin{equation}
\label{eq:varprimaEPT}
\begin{array}{ll}
\ds \delta_\epsilon \mathcal{V} & = \ds -\intop_{l_{eq}}^{\bar{l}}\left[ B\theta_{eq}^{''}(s)+ P\left(\cos\alpha\sin\theta_{eq}(s)+\sin\alpha\cos\theta_{eq}(s)\right)\right] \theta_{var}(s)ds \\ [5 mm]
&\ds +\left[B\frac{\theta_{eq}^{'}(l_{eq})^{2}}{2}- P\cos\alpha-S\right]l_{var},
\end{array}
\end{equation}
so that the equilibrium equations (\ref{system_2}) and (\ref{nonnoeshelby}) are obtained, the latter of which, representing the so-called \lq transversality condition' of Courant and Hilbert (1953), provides the Eshelby-like force.

\subsection{The Eshelby-like force expressed as a function of the transversal load}\lb{elasticazza}

The equilibrium configuration of the elastic rod satisfies the elastica equation (\ref{system_2})
 (see Love, 1927, and Bigoni, 2012) and, through
a change of variables, the rotation field (for the first mode of deformation) can be obtained as
\begin{equation}
\theta_{eq}(s)=2\arcsin\left[\eta \, \textup{sn}\left((s-l_{eq})\sqrt{\frac{P}{B}}+\mathcal{K}(m,\eta),\eta\right)\right]-\alpha,
\end{equation}
where $\textup{sn}$ is the Jacobi sine amplitude function, $\mathcal{K}\left(m,\eta\right)$ the incomplete elliptic integral
of the first kind and
\beq
\eta=\sin\frac{\theta_{\bar{l}}+\alpha}{2},\qquad
m=\arcsin\left[\frac{\sin(\alpha/2)}{\eta}\right],
\eeq
with $\theta_{\bar{l}}=\theta_{eq}(\bar{l})$ representing the rotation measured at the free end of the rod, related to the applied vertical load through
\begin{equation}\label{eq:relangolocaricoimperfiniziale}
P=\frac{B}{(\bar{l}-l_{eq})^{2}}\left[\mathcal{K}\left(\eta\right)-\mathcal{K}\left(m,\eta\right)\right]^{2}.
\end{equation}

The Eshelby-like force (\ref{nonnoeshelby}) can be expressed as
\begin{equation}
\frac{M^2}{2 B} = 2P\left(\eta^{2}-\sin^{2}\frac{\alpha}{2}\right),
\label{eq:f_conf}
\end{equation}
so that the axial force $S$ at the end of the sliding sleeve,
which will be measured through a load cell in the experiments, is given by eqn (\ref{bigonata}).
It can be noted from eqn. (\ref{bigonata}) that the
measured load $S$ is (in modulus) bounded by $P$ and that $S$ tends to $P$ only in the \lq membrane limit',
when $B$ tends to zero and $\theta_{\bar{l}} + \alpha$ to $\pi$.

The following three different cases may arise, explaining the experiments shown in Fig. \ref{fiorellino}.
\begin{itemize}
\item the elastic rod within the sliding sleeve is in compression, or \lq pushed in', if $\theta_{\bar{l}}+\alpha<\pi/2$;
\item the elastic rod within the sliding sleeve is unloaded if $\theta_{\bar{l}}+\alpha=\pi/2$;
\item the elastic rod within the sliding sleeve is in tension, or \lq pulled out', if $\theta_{\bar{l}}+\alpha>\pi/2$.
\end{itemize}

The case of null axial force, $S=0$, occurs when $M^2/(2 B)$ is equal to the
axial component of the dead load, $P\cos\alpha$, and corresponds
to deformed configurations which have the tangent at the free end orthogonal
to the direction of the dead load $P$, as in Fig. \ref{fiorellino} (center).

Finally, it can be noted that the Eshelby-like force $M^2/(2 B)$ is greater than the applied load $P$ when
\begin{equation}
\cos\alpha-2\cos^2\left(\frac{\theta_{\bar{l}}+\alpha}{2}\right)>0.
\label{eq:f_conf>P}
\end{equation}
Regions in the $\theta_{\bar{l}}-\alpha$ plane where the axial force $S$ is positive/negative and where $M^2/(2B)>P$ are shown in
 Fig. (\ref{plane}).
From the figure it can be concluded that
 $M^2/(2B)>P$ is possible only for positive axial load, $S>0$,
and high deflections of the rod (at least for rotation at the free end $\theta_{\bar{l}}$ greater than $\pi/3$ and depending on $\alpha$).
\begin{figure}[!htcb]
  \begin{center}
      \includegraphics[width= 12 cm]{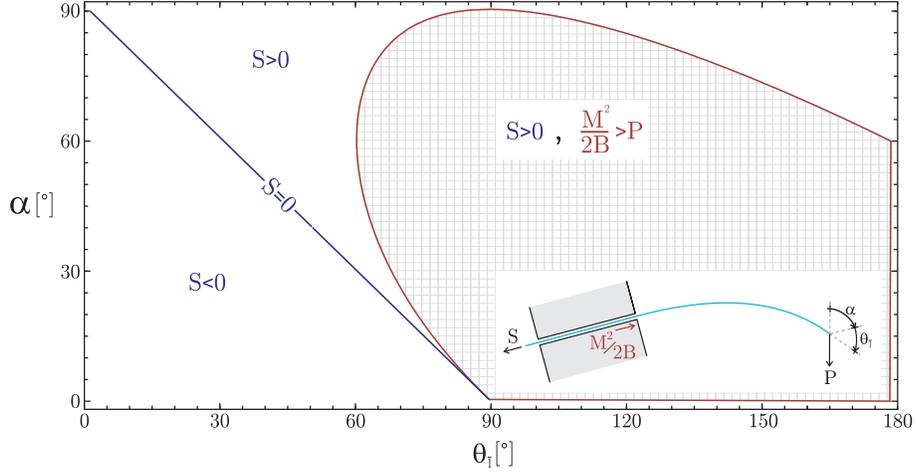}
\caption{\footnotesize Regions in the plane $\theta_{\bar{l}} - \alpha$ where $S>0$, $S<0$ and $M^2/(2 B)>P$.}
\lb{plane}
  \end{center}
\end{figure}

\section{The experimental evidence of configurational force}

The structure shown in Fig. \ref{system} has been realized using for the elastic rod two
C62 carbon-steel strips (25 mm $\times$ 2 mm cross section), one 585 mm in length and the other 800 mm. For these rods the bending stiffness $B$ has been determined with
flexure experiments to be equal to 2.70 Nm$^2$.

The sliding sleeve is 296 mm in length and has been realized with 27 pairs of rollers (made up of 10 mm diameter and 15 mm length teflon cylinders, each containing two roller bearings). The tolerance between the metal strip and the rollers is calibrated with four micrometrical screws.

The axial force $S$ has been measured using a MT1041 load cell (R.C. 300N), while dead loading, measured through a Leane XFTC301 (R.C. 500N) loading cell, has been provided with a simple hydraulic device in which water is poured at constant rate of 10 gr/s into a container. Data have been acquired with a NI CompactDAQ system,
interfaced with Labview 8.5.1 (National Instruments).
The whole apparatus has been mounted on an optical table (1HT-NM from Standa) to prevent spurious vibrations, which have been checked to remain negligible (accelerations have been found inferior to $2\times 10^{-3}$ g) with four IEPE accelerometer (PCB Piezotronics Inc., model 333B50) attached at different positions.
The tests have been performed in a controlled temperature (20$\pm$0.2 $^\circ$C) and humidity (48$\pm$0.5\%) room.
The testing set-up is shown in Fig. \ref{test_setup}. Additional material can be found in the electronic supporting material and at
http://ssmg.unitn.it/.
\begin{figure}[!htcb]
  \begin{center}
      \includegraphics[width= 13 cm]{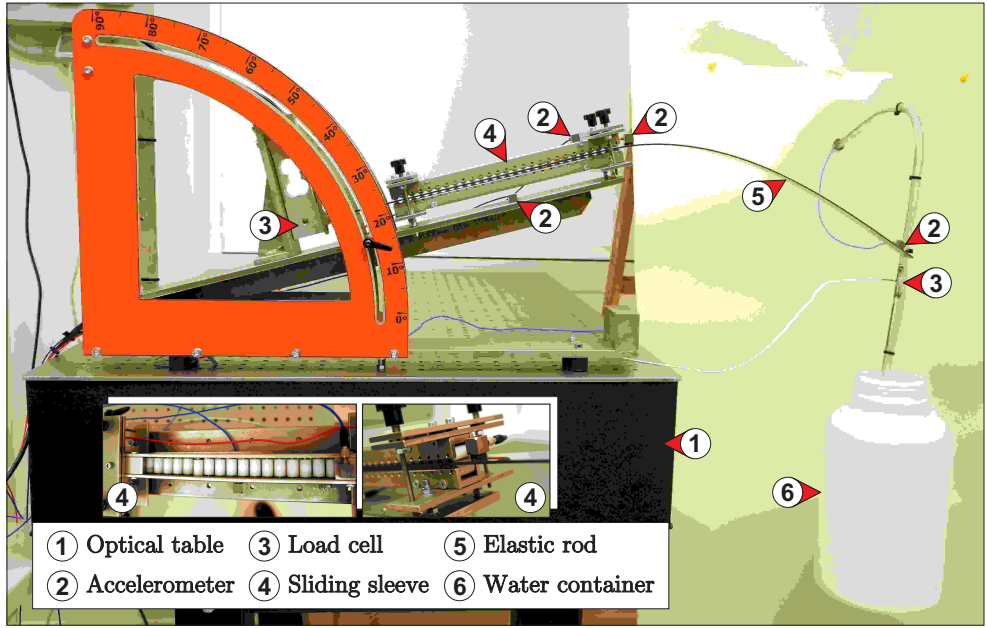}
\caption{\footnotesize The test setup for the measure of the axial Eshelby-like force transmitted by a sliding sleeve, a realization of the scheme reported in Fig. \ref{system}.}
\lb{test_setup}
  \end{center}
\end{figure}

\subsection{Eshelbian force provided by a roller device}

Rollers have been employed in the practical realization of the sliding sleeve,
so that the question may arise how this set-up is tight to our idealization and can effectively measure the Eshelby-like force.
To quantify the effects introduced by the rollers, an asymptotic approach similar to that presented
in Section \ref{asint} is developed here by considering the statically determined system given by two rollers
 with finite radius $r$ and which centers
are distant $\Delta_H+2r$
and $\Delta_V+2r$ in the axial and transversal directions,
so that the model of a perfect sliding sleeve is achieved in the limit of null value for
these three parameters ($r$, $\Delta_H$ and $\Delta_V$), Fig. \ref{roll}.
\begin{figure}[!htcb]
  \begin{center}
\includegraphics[width= 6 cm]{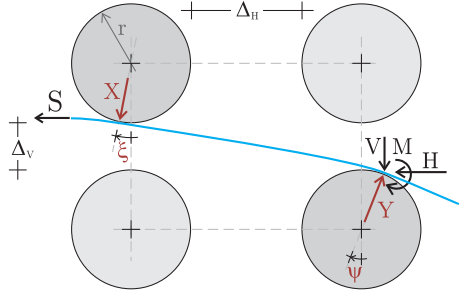}
\caption{\footnotesize The scheme of the sliding sleeve constraint realized through two pairs of rollers.}
\lb{roll}
  \end{center}
\end{figure}

In the limit $\Delta_V/ \Delta_H\rightarrow 0$, the roller reactions $X$ and $Y$ are obtained from rotational and translational (in the transversal direction) equilibrium as
\begin{equation}
X=\frac{M}{\ds\cos\xi\left[\Delta_H+r\left(2+\sin\psi+\sin\xi\right)\right]},
~~~ Y=\frac{1}{\cos\psi}\left[V+\frac{\ds M}{\ds \Delta_H+r\left(2+\sin\psi+\sin\xi\right)}\right],
\end{equation}
where $\xi$ and $\psi$ are the rotations of the rod at the contact points with the rollers,
so that the translational (in the axial direction) equilibrium leads to
\begin{equation}\label{eqorizz}
S=V\tan\psi-\frac{M\left(\tan\xi-\tan\psi\right)}{\ds\Delta_H+r\left(2+\sin\psi+\sin\xi\right)} -H.
\end{equation}
Restricting attention to small deflections between the rollers, the angles $\xi$ and $\psi$ can be obtained through
integration of the elastica as
\begin{equation}
\lb{ocamorta}
\begin{array}{l}
\ds \xi=-\frac{M (\Delta_H+2r)^{2}\left(2 B+M r\right)+6 B \left(-2 B+M r\right)\Delta_V}{2 B (\Delta_H+2r) \left(6 B + M r\right)}, \\[5mm]
\ds\psi=\frac{M (\Delta_H+2r)^{2}\left(4 B+M r\right)+6 B \left(2 B+M r\right)\Delta_V}{2 B (\Delta_H+2r) \left(6 B + M r\right)}.
\end{array}
\end{equation}

In the limit of $\Delta_V/r\rightarrow0$, eqn (\ref{ocamorta}) simplifies to
\begin{equation}
\xi=-\frac{M (\Delta_H+2r)(M r+ 4B)}{2 B (M r +6 B)}, ~~~
\psi=-\xi,
\label{eq:15}
\end{equation}
and the translational  equilibrium, eqn. (\ref{eqorizz}), reads
\begin{equation}
\lb{ocadritta}
\underbrace{\frac{M}{6 B + M r}\left[\frac{6M (3 B+ M r)}{6 B+ M r}+ \frac{V(\Delta_H+ 2 r)(4 B+ M r)}{2B}\right]}_{Eshelby-like\,\, force}
= S+H,
\end{equation}
an equation which introduces the concept of Eshelby-like force provided by a roller device, and reducing
in the limits $r\rightarrow0$ and $\Delta_H\rightarrow0$ to the value of the Eshelby-like force (\ref{nonnoeshelby}) arising from a sliding sleeve.

It can be noted that the lowest value of the configurational force realized by the roller device occurs in the limit of the sliding sleeve.

\subsection{Experiments}

Results of experiments are reported in Fig. \ref{plane2} and compared with the theoretical predictions obtained with the \lq perfect model' of sliding sleeve, eqn (\ref{nonnoeshelby}), and with the \lq roller-version' of it, eqn (\ref{ocadritta}), the latter used with parameters tailored on the experimental set up ($r=5$ mm, $\Delta_H = 1$ mm).
\begin{figure}[!htcb]
  \centering
      {\includegraphics[width= 16 cm]{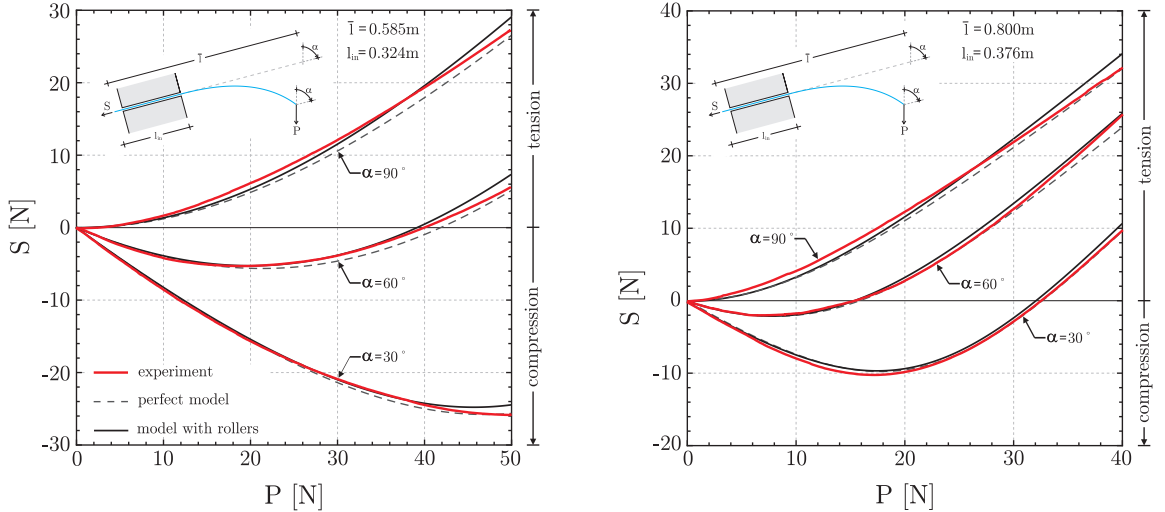}}
\caption{\footnotesize Comparison between experimental results (red curve) and the theoretical predictions. These have been reported for a perfect sliding sleeve (dashed curve) and for a sliding sleeve realized with rollers mimicking the experimental conditions (solid curve). Two rods have been used of external lengths 261 mm (left) and 424 mm (right) for different inclinations (90$^\circ$, 60$^\circ$ and 30$^\circ$).
}
\lb{plane2}
\end{figure}

First of all, we can note that the theoretical values are close to each other, which is a proof that the rollers have a negligible effect on the determination of the Eshelby-like force.
Moreover, we see that there is an excellent agreement between the theoretical predictions and the experimental results, which is an indisputable proof that Eshelby-like forces acting on elastic structures are a reality.

\section{Conclusions}

Eshelbian forces are related to the change in configuration of a mechanical system. We have shown that simple elastic structures can be designed to give evidence to these forces, that can both be calculated and experimentally detected. Therefore, the findings presented in the present article open a new perspective in the design of compliant systems.


\vspace*{5mm} \noindent
{\sl Acknowledgments } Financial support from the grant PIAP-GA-2011-286110-INTERCER2,
\lq Modelling and optimal design of ceramic structures with defects and imperfect interfaces' is gratefully acknowledged.


 { \singlespace
}

\end{document}